\documentclass[12pt]{iopart}
\usepackage[usenames,dvipsnames]{color}
\usepackage{graphicx}

\begin{document}
\title[]{Current-flux characteristics in mesoscopic  nonsuperconducting rings }

\author{L Machura$^1$, Sz Rogozinski$^1$ and J \L uczka$^1$}

\address{$^1$ Institute of Physics, 
University of Silesia, Katowice, Poland}
\ead{lukasz.machura@us.edu.pl}

\begin{abstract}
We propose four different mechanisms responsible for paramagnetic or diamagnetic  persistent currents in normal metal rings and determine the circumstances for change of the  current 
 from paramagnetic to diamagnetic ones and {\it vice versa}.   It might qualitatively 
 reproduce the experimental results of  Bluhm et al. (Phys. Rev. Lett. 102, 136802 (2009)).
\end{abstract}

\submitto{\JPCM}
%\maketitle

\section{Introduction}

In the absence of an applied voltage, an induced electrical current rapidly decays due  to dissipation processes. In  normal  (not superconducting) metal rings  it  typically dies out within the relaxation time of 
order  $ 10^{-13} $ s. However, if a radius  of the ring is small
enough (below  microns) and   temperature of  the system is 
below $1$ K,   quantum effects  start to play a distinct role. Under right  circumstances electrons in the ring 
are able to preserve its coherence which in turn results in the persistent  (dissipationless) current 
induced by the static applied magnetic field. 

The existence of  persistent currents in  metallic rings was predicted by Hund in 1938
\cite{hund}. More than 30 years later Bloch \cite{bloch} and Kulik \cite{kulik} confirmed this
prediction by  means of  quantum--mechanical models. The  strong  interest in   physics of  
mesoscopic rings arose after the 1983 paper   \cite{but} where the authors showed  that persistent currents could  flow even in the presence of disorder. 
      Experiments on persistent currents have produced a number of confusing results in apparent contradiction with theory and even amongst the experiments themselves (e.g. the response of rings was 10-200 larger than  theoretically predicted) \cite{chandra}.   We could observe persistent controversy on persistent currents  for nearly twenty years. 
  Recently, two  groups \cite{bluhm,harris} have developed new different techniques which  allow  to make measurements a full order of magnitude more precise than any previous attempts.    Both  experiments  confirm to a high degree  the physics theory regarding the behavior of persistent currents. 
The  group of K. Moler   \cite{bluhm} has  employed a scanning 
technique (a SQUID microscope) and  measured the magnetic response of 33 individual  mesoscopic gold rings.   Each ring was scanned individually, unlike past experiments on persistent currents conducted by other groups. In total  the rings  were scanned approximately 10 million times.
The  team  of J. Harris \cite{harris} has  developed an  alternative measurement scheme.  
The  aluminum rings have been deposited on a cantilever used as a torque magnetometer   whose vibration frequency can be precisely monitored. From the frequency shift caused by the magnetic flux, the researchers could deduce the current with a precision of two orders of magnitude greater than it was  possible in the past. 
They have  studied several different cantilevers decorated with a single aluminium ring or arrays of hundreds or thousands of identical aluminium rings. The rings on different cantilevers had  radiuses  of 308-793 nanometers. 

 Persistent  currents are  highly sensitive to a variety of  factors.    It is clearly  visible in experimental data shown  in  panel (b)  of  Fig. 2 in  Ref. \cite{bluhm}: for nominally identical samples the observed response  is  paramagnetic (e.g. for the ring 1) or  diamagnetic (e.g. for the ring 2). 
Here, we propose several  possible mechanisms
for controlling the  response  of the metallic mesoscopic rings and determine the conditions
for which the transition from paramagnetic to diamagnetic current is able to occur. 

The paper is  arranged as  follows.  In the
section \ref{2fm}, we start with the presentation of the model  for  the flux dynamics in the rings. 
Next, in the section \ref{results}, we identify the operating conditions on which 
similar rings  can exhibit  opposite responses. We finalize the paper with conclusions in the section 4.  

%%%%%%%%%%%%%%%%%%%%%%%%%%%%%%%%%%%%%%%%%%%%%%%%
\section{Model  for  flux dynamics of  mesoscopic rings}\label{2fm}

At  low temperature, small  normal  metal  rings threaded by a magnetic flux can display 
persistent and non-dissipative  currents  carried by  phase-coherent
electrons.  The circumference of the ring should be smaller than the electron’s phase coherence length.  This typically limits the sample size to below  micrometers and the temperature to below 1 K. 
However, at  temperature $T>0$,  a part of  electrons  looses 
its phase-coherence due to  thermal fluctuations and constitutes a dissipative Ohmic current 
associated with the resistance $R$. 
The actual magnetic flux $\phi$ induced  by the current flowing in the  ring is given by 
the relation 
\begin{equation}\label{LL1}
\phi =\phi_e + L I,  
\end{equation}
where $ \phi_e $ is the magnetic flux generated by the external constant magnetic field, $ I $ denotes the total current 
flowing in the ring and $ L $ stands for the self--inductance of the ring. Dynamics of the magnetic flux 
$\phi$ in such a system is modeled  by the dimensionless Langevin-type equation \cite{rogo} 
\begin{equation}\label{QOV}
\frac{dx}{ds} = - \frac{dV(x)}{dx} + \sqrt{2D_{ \lambda}(x)}\;\Gamma (t), 
\end{equation}
where  $x=\phi/\phi_0$ is the rescaled magnetic flux  and $\phi_0=h/2e$ is the flux quantum.  The rescaled time 
$s=t/\tau_0$, where the characteristic time $\tau_0=L/R$    is    the inductive  time of the ring. For a typical mesoscopic ring, $L/R$  is in the picosecond range.  The function 
\begin{eqnarray} \label{V(x)}
V(x)=\frac{1}{2}(x-x_e)^2 + B(x),
\end{eqnarray}
where  $x_e=\phi_e/\phi_0$ and \cite{rogo}
\begin{eqnarray}
  \label{B}
 B(x)= \alpha \sum_{n=1}^{\infty}\frac{A_n(T_0)}{2n\pi} \cos(2n\pi x) [p + (-1)^n (1-p)], 
\end{eqnarray}
where $\alpha = LI_0/\phi_0$ and  $I_0$ is the maximal  persistent current at zero temperature.
In the mechanical context, $V(x)$   could be interpreted as an effective  potential. 
 The temperature dependent amplitudes $A_n(T_0)$ are determined by the relation 
\cite{cheng}
\begin{eqnarray}
A_n(T_0)= \frac{4T_0}{\pi}\frac{\exp(-nT_0)}{1-\exp(-2nT_0)}
\end{eqnarray}
with the dimensionless temperature $T_0=T/T^*$,  
 where the  characteristic temperature $T^*$ is  proportional to the
energy gap $\Delta _F$ at the Fermi surface. 
The persistent current strongly depends on the parity of the number of  coherent electrons. It is  taken into account by  assigning the 
probability $p$ of an even number of the coherent electrons. Then the corresponding probability
of an odd number of the coherent electrons is given by $ 1 - p $. 

Thermal equilibrium fluctuations are modeled  by  $ \delta $--correlated Gaussian white noise  $\Gamma(t)$ 
of zero mean. Classically this situation holds true in many cases. When temperature is lowered, however, the
quantum nature of thermal  fluctuations becomes important and  starts to play a role.  Therefore 
the  standard diffusion coefficient $D_0=k_BT/R$  ($k_B$
denotes the Boltzmann constant)  is  modified due to quantum effects like tunnelling, quantum reflections and purely quantum fluctuations \cite{ankerhold1,luczka,coffey}.  
The modified diffusion coefficient $ D_\lambda $  assumes the form  \cite{luczka,rogo}
\begin{eqnarray}
\label{D(x)}
D_{\lambda}(x)=\frac{\beta^{-1}} {1-\lambda\beta V''(x)}   
\end{eqnarray}
with  $\beta^{-1}= k_BT/2E_m = k_0 T_0$,   the 
elementary magnetic flux energy $E_m=\phi_0^2/2L$ and 
$k_0= k_BT^*/2E_m$ is the ratio of two characteristic energies.  The  prime denotes differentiation with respect to $x$.   
The dimensionless quantum correction parameter  \cite{ankerhold1}
\begin{equation}\label{lam}
\lambda =  \lambda_0 \left[ \gamma + 
\Psi\left(1+   \frac{\epsilon}{T_0}\right)  \right],
\quad
\lambda_0=\frac{ \hbar R}{\pi\phi_0^2}, 
\quad
\epsilon = \frac{\hbar/2\pi CR}{k_BT^*},
\end{equation}
where the psi function $\Psi(z)$ is the logarithmic derivative of the Gamma function,  $\gamma \simeq 0.5772$ 
is the Euler gamma constant and $C$ is capacitance of the system related to charging effects.  
The  parameter $ \lambda $ characterizes quantum corrections to classical thermal fluctuations and  can be 
formulated as the difference between  quantum and classical fluctuations of the dimensionless flux,   
\begin{equation}\label{Lam1}
\lambda=\langle x^2\rangle_{q} -\langle x^2\rangle_{c} 
\end{equation}
where   $\langle \cdot \rangle$ denotes thermal equilibrium average,  the subscripts 
$q$ and $c$ refer to quantum and classical cases, respectively.  
Let us  remember  that  the diffusion coefficient $ D_\lambda(x) $ cannot be negative and therefore  the 
parameter $ \lambda $ has to be chosen small enough to ensure its non-negativeness  for any argument.
 Because in Eq. (\ref{QOV}) the noise term contains the multiplicative white noise  $\Gamma(t)$,  it is important to stress that Eq. (\ref{QOV})   has to be interpreted  in the Ito sense \cite{gard}.  Therefore 
the corresponding Fokker--Planck equation for the time evolution of the probability density $ P(x, t) $
reads \cite{gard}
\begin{equation}\label{FP}
\frac{\partial}{\partial t} P(x, t) = \frac{\partial}{\partial x} \left[
\frac{dV(x)}{dx} P(x, t) \right] + \frac{\partial^2}{\partial x^2} 
 \left[ D_{\lambda}(x) P(x, t) \right] . 
\end{equation}
The average stationary dimensionless current $i$ flowing in the ring can be calculated from Eq. (\ref{LL1}): 
\begin{equation} \label{Ia}
i = \langle x \rangle - x_e, \quad i= \langle I \rangle L/\phi_0 
\end{equation}
and the average stationary magnetic flux $ \langle x \rangle$ is  calculated from the relation  
\begin{eqnarray} 
\label{phia}
 \langle x \rangle = \int_{-\infty}^{\infty} x \; P(x) d x,
\end{eqnarray}
where $P(x)=\lim_{t\to \infty} P(x, t)$ is a stationary probability density being a solution of the Fokker-Planck equation (\ref{FP})  for   $\partial P(x, t)/\partial t =0$ and  zero  stationary probability current. It has the form 
\begin{eqnarray}\label{ps}
P(x)= N_0  D^{-1}_{\lambda}(x) \exp\left[-\Psi_{\lambda}(x)\right], 
\end{eqnarray}
where $N_0$ is the normalization constant and 
the generalized thermodynamic potential  $\Psi_{\lambda}(x)$  reads
\begin{eqnarray} 
\label{Psi}
\Psi_{\lambda}(x) =\int\frac{dV(x)}{dx} \,D^{-1}_{\lambda}(x) \,dx 
= \beta V(x) - \frac{1}{2} \lambda \beta^2 \left[ V'(x) \right]^2.
\end{eqnarray}
In the case when thermal fluctuations can be treated classically, i.e.  when $ \lambda =0$, the stationary state $P(x)$  is described by the Boltzmann distribution.  When thermal fluctuations have to be considered as quantum fluctuations, the stationary state is still a thermal equilibrium state but now described by   the non-Boltzmann distribution (\ref{ps}) with  the $ x $-dependence of the diffusion coefficient  $ D_\lambda(x) $.

\section{Current--flux characteristics}\label{results}

We are interested in the current-flux characteristics, i.e. in dependence of the stationary current $i=i(x_e)$ on the applied magnetic flux $x_e$.  To this aim we exploit  Eqs. (\ref{Ia})-(\ref{Psi}).  Note that the external flux  $x_e$ enters   Eq. (\ref{Ia})  and the remaining  equations  via the potential (\ref{V(x)}) which occurs both in the diffusion function $D_{\lambda}(x)$ and the generalized thermodynamical potential $\Psi_{\lambda}(x)$.   Because the persistent current  is a periodic function of the  magnetic flux, we consider the current-flux characteristics on the unit interval of $x_e$. 
We present  four visually similar sets of the current-flux characteristics  with various  set-ups to demonstrate sensitivity of the persistent currents to some subtle effects. It could  explain 
 the different response  of  the nominally identical metal rings to the magnetic flux  presented in the recent experimental
work \cite{bluhm}.   The measured  persistent current of 15 nominally identical rings with radius 
$R=0.67 \mu$m exhibits both the paramagnetic and diamagnetic response. 
 In  panel (b) of  Fig.  2 in Ref.\cite{bluhm},  one can easily notice that in the linear 
response regime, i.e.  for  values of the external flux close to zero,  the  susceptibility defined as the 
ratio of the average persistent current  to   the external flux has sometimes positive slope indicating the 
 paramagnetic current (see   the current for the rings  1, 3, 4, 10, 11, 12, 15 in Fig.  2(b) in \cite{bluhm}) and sometimes negative slope 
exhibiting the  diamagnetic response (see the current for the  rings 2, 5, 6, 7, 8, 9, 13, 14 in Fig.  2(b) in \cite{bluhm}). 

\begin{figure}[tbp]
\includegraphics[width=0.99\linewidth]{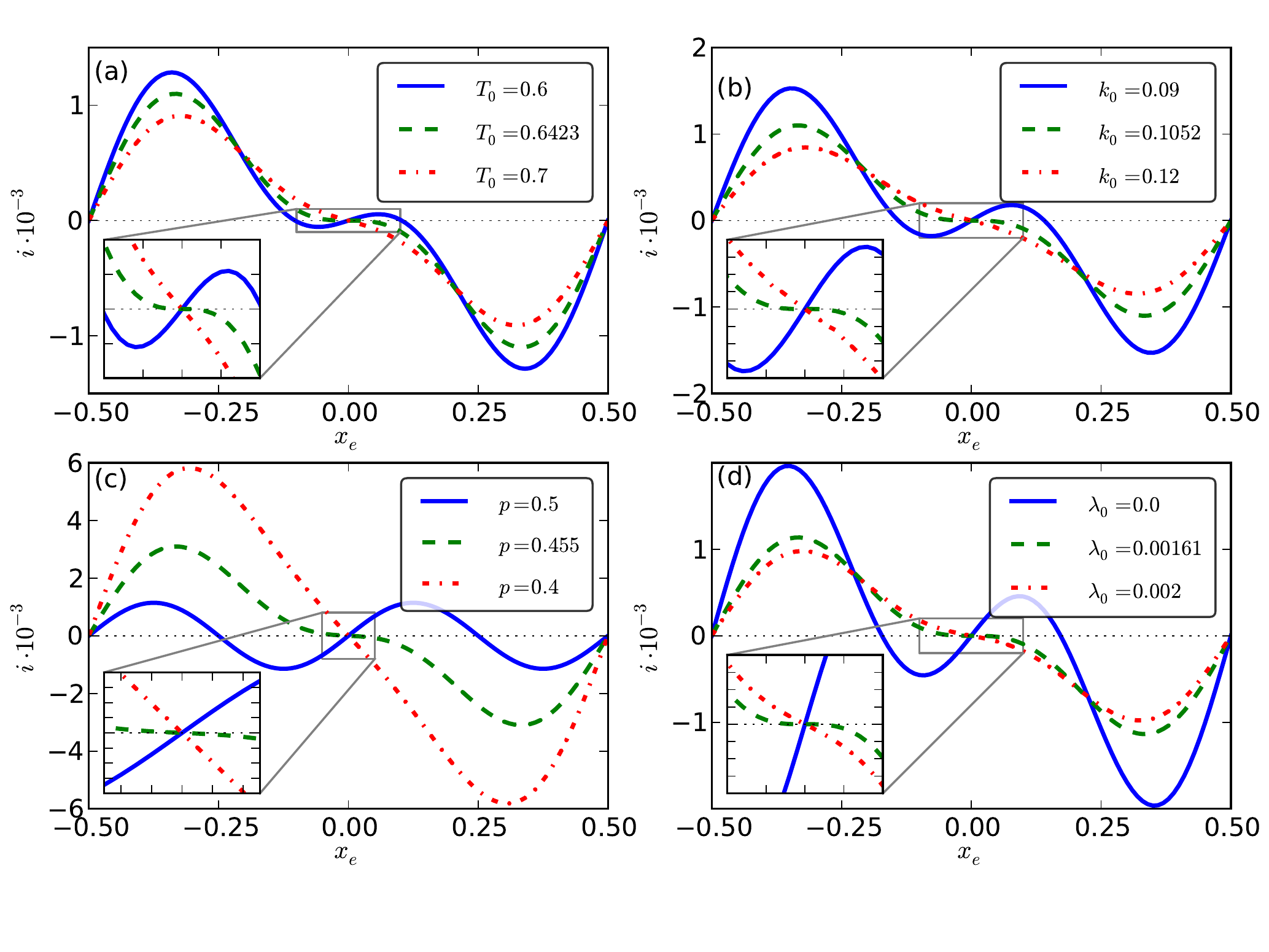} 
\caption{(color online) The stationary averaged velocity {\it vs} the external magnetic flux $ x_e $.
In the panel (a) we show current--flux characteristics in the classical regime for three different 
temperatures $ T_0 = 0.6, 0.6423, 0.7 $.
Panel (b) presents also the classical regime for three different structure constants
$ k_0 = 0.09, 0.1052, 0.12 $.
Panel (c) exhibits again the persistent current for ring working in the classical regime for three 
values of the probability $ p = 0.4, 0.455, 0.5$ of an even number of coherent electrons in the ring.
Finally panel (d) reveals the influence of the quantum parameter $ \lambda_0 = 0$ (indicating classical 
regime), $ 0.00161, 0.002 $ (quantum regime).
Blue (solid) lines mark the paramagnetic response to the external stimulus, green (dashed) lines denote
the situation where the magnetic susceptibility is zero and finally red (dashed--dotted) lines indicate
the diamagnetic susceptibility of the normal metal mesoring around zero external magnetic flux
(see insets for details). 
The parameters not given explicitly are set as follows: $ T=0.5 $, $ p = 0.48 $, $ k = 0.08 $, 
$ \alpha = 0.1 $.
}
\label{fig1}
\end{figure}

Figure \ref{fig1}  illustrate  possible current-flux characteristics.   
All parameters are chosen arbitrarily  as we don't aim to compare our results with the experimental data.  The goal of this work is rather to constitute possible mechanisms  responsible 
for  diamagnetic or  paramagnetic responses. Neverthless,  all curves corresponding to the paramagnetic currents (blue solid lines) in all
panels look very similar to the experimental curve shown in figure S6(A) in the Supporting Online Material 
\cite{supp} of the paper \cite{harris}. 

In the panel (a) we depict the current--flux characteristics in the regime of classical thermal fluctuations (i.e. when $\lambda =0$)  for three different 
temperatures.  For  the lowest  temperature $ T_0 = 0.6 $ (blue solid line) the situation 
reveals the paramagnetic response of the ring to the applied magnetic flux. For the higher  
temperature $ T_0 = 0.7 $ (red dashed--dotted line) the response  is diamagnetic. 
The cross--temperature below which the response is  paramagnetic and above which is diamagnetic was recognized 
numerically at the value of $ T_0^C = 0.6423 $ (see green dashed line and the inset of the panel (a) for details).

Panel (b) presents the ring reaction to the external stimulus for  three  values of the 
structure parameter $ k_0= k_B T^* / 2 E_m $. It stands for the ratio of the two characteristic energies, the thermal energy  $k_B T^* / 2$ to  the magnetic energy  $ E_m = \phi_0^2 / 2 L $.  
 It is  instructive to define
$ k_0 $ in the  alternative way as  $ k_0 = L I_0 / \phi_0 $. At the value of $ k_0 = 0.09 $ the response is paramagnetic (blue
solid line), for $ k_0 = 0.12 $ it is diamagnetic (red dashed-dotted line) and again we have numerically 
identified the cross--value at the level of $ k_0^C = 0.1052 $ (green dashed line).  We note that 
the ratio of  two parameters $ r_{\alpha} = \alpha / k_0 = i_0 / I_0 \sim l_e / l$ describes the physical properties 
of the metal ring \cite{book}. Here $ l_e $ is the elastic mean free path of the electron in the ring and $ l $ stands 
for the circumference of the ring. For the multichannel rings the above ratio is greater than one for the 
ballistic regime and smaller than one in the diffusive one. In the presented collection of the parameters
the cross--value of the para-- to -- diamagnetic response lies slightly below the value of unity, i.e. at 
$ r_\alpha^C \simeq 0.95$.  It  means that this border value lies in the diffusive regime and all rings 
designated by the higher values of $ k_0 $ showing the diamagnetic susceptibility will also lie in the 
same diffusive regime. 

It is well know that the most straightforward way of turning the susceptibility from diamagnetic to paramagnetic is to change
the probability $ p $ of an even number of coherent electrons in the ring. In the panel (c) we present 
the current for three different rings with probabilities $ p = 0.5$ (blue solid line), $0.455$ (green
dashed line) and $0.5$ (red dashed--dotted line) exhibiting the paramagnetic, zero and diamagnetic 
susceptibility, respectively. 

In the last panel (d) we focus on the influence of quantumness of  thermal fluctuations on persistent currents.
The comparison of the classical thermal fluctuations with $ \lambda_0 = 0 $ and corresponding paramagnetic current
(blue solid line) with two values of the non--zero quantum noise parameter $ \lambda_0^C = 0.00161 $ 
exhibiting zero magnetic susceptibility (green dashed line) and $ \lambda_0 = 0.002 $ with diamagnetic 
response (red dashed--dotted line) is presented. As shown, the sign of the current in the vicinity of
zero magnetic field can be easily affected by small perturbation of the quantum correction parameter 
$\lambda$.

\section{Conclusions}
We have proposed four  various  mechanisms that can lead to either diamagnetic or paramagnetic currents 
in  the metal rings. Two of them are  related to  the physical properties of the metal
rings. These are the structure parameter $ k_0 $ together with the probability $ p $ of an even number 
of coherent electrons in the ring. Another two can be adjusted by tuning the temperature of the system.
In the experiment \cite{harris}, the temperature uncertainty is $7\%$ \cite{supp}. 
In all  presented cases,  the small change of the control parameter causes the reversal of the susceptibility - from paramagnetic to diamagnetic ones or {\it vice versa}. 
In the nowadays experiments on the mesoscopic rings it is impossible to justify which system parameters 
are responsible for the observed responses. The fact that the scientists are able to perform 
the measurements on single separated rings is a great success and a milestone in the present 
state--of--the--art. In the near future the experimental physicists will be able to prepare experiments in 
the desired and more precise conditions and then  our  findings  might  be tested.  

\ack
The work supported by the ESF Program {\it Exploring the Physics of Small Devices}.

\end{document}